\documentclass[10pt]{article}
\usepackage{amsmath}
\usepackage{mathrsfs}
\usepackage[dvipdfmx]{graphics}
\usepackage{hyperref}
\usepackage[dviout]{graphicx}
\usepackage[at]{easylist}
\usepackage{ascmac}
\usepackage{physics}
\usepackage{here}
\usepackage{cite}
\usepackage{amssymb}
\usepackage{mhchem}
\usepackage{authblk}
\usepackage{indentfirst}
\usepackage[subrefformat=parens]{subcaption}
\usepackage{color}

\usepackage[top=30truemm,bottom=30truemm,left=25truemm,right=25truemm]{geometry}
\makeatletter
\@addtoreset{equation}{section}

\makeatother
\definecolor{DarkBlue}{rgb}{0,0,0.9} 

\definecolor{DarkRed}{rgb}{0.65,0,0}

\title{False Vacuum Decay in Rotating
BTZ 
Spacetimes
}
\date{ }
\author{Daiki Saito\footnote{Email:saito.daiki.g3@s.mail.nagoya-u.ac.jp}}
\author{Chul-Moon Yoo\footnote{Email:yoo@gravity.phys.nagoya-u.ac.jp}}
\affil{Division of Particle and Astrophysical Science, Graduate School of Science, Nagoya University, Nagoya 464-8602, Japan}

\begin{document}

\maketitle

\begin{abstract}
We analyse vacuum decay in rotating BTZ black hole spacetimes with the thin wall approximation.
Possible parameter regions for the vacuum decay are clarified. 
We find that the nucleation rate is dominated by the bounce solution with the static shell configuration. 
The nucleation rate of the static shell decreases with the mass of the initial black hole. 
For a larger/smaller value of the initial black hole, 
the nucleation rate can be smaller/larger than 
that of the Coleman De Luccia vacuum decay in the pure AdS spacetime. 
Through the vacuum decay, the black hole gains its mass and loses the horizon area. 
We also find that the nucleation rate increases with increasing the angular momentum of the spacetime. 
\end{abstract}

\tableofcontents

\section{Introduction}

Vacuum decay is a transition from a classically stable state (false vacuum) 
to a lower energy state (true vacuum),
which may be described by a quantum tunneling process.  
By considering this transition in the spacetime occupied by a false vacuum state, 
a true vacuum region surrounded by a bubble wall can be realized  
due to the quantum tunnelling. 
The nucleated bubble typically 
expands, and eventually, the spacetime will be filled with the true vacuum. 
This phenomenon was considered for the first time by Coleman over 40 years ago \cite{Coleman:1977py,Callan:1977pt}. 
Since gravity universally couples with other fields,  
it would be important to take into account 
the effect of gravity in vacuum decay. 
Pioneering work has been done 
by Coleman and De Luccia in Ref.~\cite{Coleman:1980aw},
where the decay in 
a maximally symmetric spacetime 
was discussed. 
Vacuum decay in a black hole (BH) spacetime was firstly considered in Ref.~\cite{Hiscock:1987hn}, 
which stated that the four-dimensional Schwarzschild BH 
acts as a nucleation site for the decay, and more sophisticated researches have been performed in Refs.~\cite{Gregory:2013hja,Burda:2015yfa}. 
In particular, 
the authors in Refs.~\cite{Gregory:2013hja,Burda:2015yfa} 
considered the cases in which 
the BH mass changes through the transition associated with the bubble nucleation, 
and reported that the mass and the horizon area of the BH may decrease in some cases.

As far as we know, the physical meaning of the vacuum decay in BH spacetimes has not yet been sufficiently clear. 
One plausible interpretation of the effect of a BH on the vacuum decay would be a thermal environmental system \cite{Mukaida:2017bgd}. 
In Ref.~\cite{Mukaida:2017bgd}, it is
stated that the vacuum decay in the Schwarzschild (de-Sitter) spacetime can be regarded as 
the activated tunnelling in the flat background with the finite temperature equal to the Hawking temperature of the BH. 
That is, the thermal radiation from the black hole supports the vacuum phase transition.  
On the basis of this interpretation, 
the effect of the black hole on the vacuum decay may involve the quantum nature of gravity. 
The decreasing horizon area reported in Ref.~~\cite{Gregory:2013hja,Burda:2015yfa} 
means the violation of the black hole area theorem 
in the classical 
theory and might also
indicate
the quantum gravity effect. 
Therefore, 
it might be expected that 
the vacuum decay in BH spacetimes
is a clue to understanding the quantum nature of gravity. 

Vacuum decay in BH spacetimes is also attractive from the aspect of 
several
applications. 
The Higgs particle 
has been discovered in 
the LHC experiment in 2012 \cite{ATLAS:2012yve,CMS:2012qbp}, 
and our universe might be
in a false vacuum of Higgs potential \cite{Sher:1988mj,PhysRevD.40.613,Espinosa:1995se,Isidori:2001bm,Elias-Miro:2011sqh,Degrassi:2012ry,Buttazzo:2013uya}. 
The estimated lifetime of the false vacuum in the flat background spacetime 
is longer than the age of the universe \cite{Chigusa:2017dux,Chigusa:2018uuj}. 
However, the lifetime can be 
decreased by the existence of BHs in the universe as is reported in Refs.~\cite{Burda:2015isa,Burda:2016mou,Gregory:2016xix,Kohri:2017ybt}. 
Based on this result, 
by considering the effects of primordial black holes (PBHs) on the vacuum decay, 
Ref.~\cite{Dai:2019eei} set constraints on the number and mass 
of PBHs.

From the aspect of the application, we should 
note that seed BHs are restricted to static and spherically symmetric ones in
most researches. 
However, in general, BHs have 
non-zero values of the angular momentum, 
and the analyses with
static and spherically symmetric solutions would not be sufficient for the application in our universe.  
In order to get a deeper insight into the fundamental understanding and general property of the vacuum decay in BH spacetimes, 
it would be helpful to collect more variety of vacuum decay phenomena in BH spacetimes.
In particular, in this paper, we 
focus  
on the effects of 
the angular momentum. 
In four-dimensional spacetimes, however, 
the analyses of vacuum decay without spherical symmetry become significantly difficult because the bubble dynamics cannot be reduced to 
one-dimensional particle mechanics as in Ref.~\cite{Gregory:2013hja}. 
No formal procedure is known for treating the bubble nucleation in a non-spherical BH spacetime, and 
there  
can be found 
only one 
attempt in Ref.~\cite{Oshita:2019jan} with 
some assumptions. 
In this paper, 
in order to see effects of the angular momentum without the difficulty,
we consider the vacuum decay 
in rotating BTZ BH spacetimes, 
which 
are three-dimensional asymptotically AdS spacetimes. 
In the 
three-dimensional asymptotically AdS spacetime, 
the BH spin does not violate 
the spherical symmetry, and 
we can proceed in the same way as Ref.~\cite{Gregory:2013hja}.

This paper is organized as follows.
In Sec.~\ref{sec2}, we see the Euclidean metric of the BTZ spacetime and 
consider the 
matching conditions for two BTZ spacetimes separated by a thin spherical shell. 
The shell equations of motion are derived in Sec.~\ref{sec2.3} and we
show that 
the shell dynamics is reduced 
to a one-dimensional potential problem of  
particle dynamics. 
In Sec.~\ref{sec3}, we 
clarify the possible parameter regions for the bubble nucleation. 
In Sec.~\ref{sec4}, we 
derive the explicit expression of the 
vacuum decay rate and show 
the results in Sec.~\ref{sec5}. 
Sec.~\ref{sec6} is devoted to a summary and conclusion.

Throughout this paper, we use the geometrized units in which both the speed of light and the Newton's gravitational constant are unity, $G=c=1$.
\footnote{Note that many papers treat BTZ spacetime with BTZ unit, $8G=c=1$.} 

\section{Geometry of the Shell and Equations of Motion}
\label{sec2}

\subsection{Lorentzian and Euclidean Metric}
\label{sec2.1}

The line element in the BTZ spacetime is given 
as follows in the Boyer-Lindquist coordinates \cite{PhysRevLett.69.1849,PhysRevD.48.1506}:
\begin{align}
ds^2&=-f(r)dt^2+\frac{1}{f(r)}dr^2+r^2\left(d\varphi-\frac{4J}{r^2}dt\right)^2 \label{eq:ds} 
\end{align}
with
\begin{align}
f(r)&=-8M+\frac{r^2}{l^2}+\frac{16J^2}{r^2}. 
\end{align}
Here, $M$ and $J$ are the mass and the angular momentum of the BH, 
respectively, and 
$l$ is the AdS length, which is 
given by $l^{-2}=-\Lambda$ with $\Lambda$ being the cosmological constant. 

In order to evaluate the decay rate, 
we need to calculate geometric quantities in the Euclidean spacetime. 
We can get the Euclidean metric 
through the transformations $t=-it_{E}$ and $J=-iJ_{E}$:
\begin{align}
ds^2&=f_{E}(r)dt_{E}^2+\frac{1}{f_{E}(r)}dr^2+r^2\left(d\varphi+\frac{4J_{E}}{r^2}dt_{E}\right)^2,  \label{eq:dsE} 
\end{align}
where
\begin{align}
f_{E}(r)&=-8M+\frac{r^2}{l^2}-\frac{16J_{E}^2}{r^2}. 
\end{align}

This paper focuses on vacuum decay in 
%a 
the BTZ spacetime with 
the thin wall approximation. 
We assume that the spacetime is spherically symmetric, and therefore the spacetime after the bubble nucleation is also described by a BTZ spacetime.
Since the shape of the bubble is also spherical, we can express the bubble wall trajectory as $r=R(t)$. 
Then, for convenience, we
define a frame that is co-rotating with the spacetime on the bubble wall, that is, the non-diagonal component 
of the metric vanishes on the bubble wall.  
Following Ref.~\cite{Lemos:2015xwa}, 
we transform the azimuthal coordinate $\varphi$ to $\phi$ defined by
\begin{align}
d\phi&:=\left.\left(d\varphi+\frac{g_{t\varphi}}{g_{\varphi\varphi}}dt\right)\right|_{r=R(t)}\nonumber \\
&=d\varphi-\frac{4J}{R^2(t)}dt, \label{eq:corot}
\end{align}
and rewrite the metric as follows:
\begin{align}
ds^2&=\left\{-f(r)+16r^2J^2\left(\frac{1}{R^2(t)}-\frac{1}{r^2}\right)^2\right\}dt^2+\frac{1}{f(r)}dr^2+r^2d\phi^2+8Jr^2\left(\frac{1}{R^2(t)}-\frac{1}{r^2}\right)dtd\phi \label{eq:dsN} \\
&=\left\{f_{E}(r)+16r^2J_{E}^2\left(\frac{1}{R^2(t_{E})}-\frac{1}{r^2}\right)^2\right\}dt_{E}^2+\frac{1}{f_{E}(r)}dr^2+r^2d\phi^2-8J_Er^2\left(\frac{1}{R^2(t)}-\frac{1}{r^2}\right)dt_{E}d\phi.
\end{align}

\subsection{Bubble Wall as a Hypersurface}
\label{sec2.2}

The bubble wall is given as a hypersurface $\mathcal{W}:r-R(t_{E})=0$
whose unit normal vector $n_{\mu}$ is 
given by
\begin{align}
n_{\mu}&=C_{E}(-\partial_{t_{E}}R,1,0) 
\end{align}
with 
\begin{align}
C_{E}&=\left[\frac{1}{f_{E}(R)}(\partial_{t_{E}}R)^2+f_{E}(R)\right]^{-1/2} 
\end{align}
in the coordinate bases associated with $(t_{E},r,\phi)$. 
We use $\phi$ and the proper time $\tau$ of the radial observer on the bubble wall as the intrinsic coordinates of the wall.
The four-velocity of the radial observer can be written as
\begin{align}
v^{\mu}&=\left(\dot{t}_{E},\dot{R},0\right), 
\end{align}
where
the dot denotes the derivative 
with respect to 
$\tau$. 
From 
the normalization condition of the four velocity $g_{E\mu\nu}v^{\mu}v^{\nu}|_{r=R(t)}=1$, 
we obtain
\begin{align}
f_{E}(R)\dot{t}_{E}=\sqrt{f_{E}(R)-\dot{R}^2}. \label{eq:AB}
\end{align}
From this Eq.~\eqref{eq:AB}, we obtain $C_{E}=\dot{t}_{E}$, and can rewrite the normal vector as
\begin{align}
n_{\mu}&=\left(-\dot{R},\dot{t}_{E},0\right). \label{eq:nL} 
\end{align}
The components of the projection tensor onto $\mathcal{W}$ are
given by 
\begin{align}
e^{\mu}_{\tau}&=v^{\mu}=\left(\dot{t}_{E},\dot{R},0\right), \label{eq:et} \\
e^{\mu}_{\phi}&=(0,0,1) .  \label{eq:ep}
\end{align}

\subsection{Junction Conditions}
\label{sec2.3}

In this subsection, 
we derive 
equations of the shell motion 
from the Israel's junction condition \cite{Israel:1966rt}. 
We label 
physical quantities on the spacetime before/after the nucleation with the subscript $+$/$-$.  
The junction condition consists of the 1st junction condition
\begin{align}
[h_{ab}]_{\pm}=0 \label{eq:1st}
\end{align}
and the 2nd junction condition
\begin{align}
[K_{ab}]_{\pm}=-8\pi\left(S_{ab}-h_{ab}S\right), \label{eq:2nd}
\end{align}
where, for convenience,  
we used the expression 
\begin{align}
[A]_{\pm}:=A_{+}-A_{-}, 
\end{align}
and $S_{ab}$ is the energy-momentum tensor on the shell. 
$h_{ab}$ and $K_{ab}$ are the induced metric and the extrinsic curvature on the shell, respectively. 
In the Euclidean spacetime, they are defined 
as
\begin{align}
h_{Eab}&:=e^{\mu}_{a}e^{\nu}_{b}g_{E\mu\nu}, \label{eq:h} \\
K_{Eab}&:=e^{\mu}_{a}e^{\nu}_{b}\nabla_{\mu}n_{\nu}. \label{eq:K}
\end{align}
In this paper, we assume that the energy-momentum on the shell is given as follows:
\begin{align}
S_{Eab}=-\sigma h_{Eab} \label{eq:Stension}
\end{align}
where $\sigma$ is the value of the tension. 
In the current situation, the induced metric is
\begin{align}
h_{E\pm ab}
=
\left(
\begin{array}{cc}
1 & 0 \\
0 & R^2
\end{array}
\right)
\end{align}
and Eq.~\eqref{eq:1st} is automatically satisfied. 
From 
Eq.~\eqref{eq:Stension}, we can rewrite Eq.~\eqref{eq:2nd} as
\begin{align}
[K_{Eab}]_{\pm}=-8\pi\sigma h_{Eab}. \label{eq:2nd'}
\end{align}

\begin{figure}[H]
\begin{center}
  \includegraphics[clip,width=6cm]{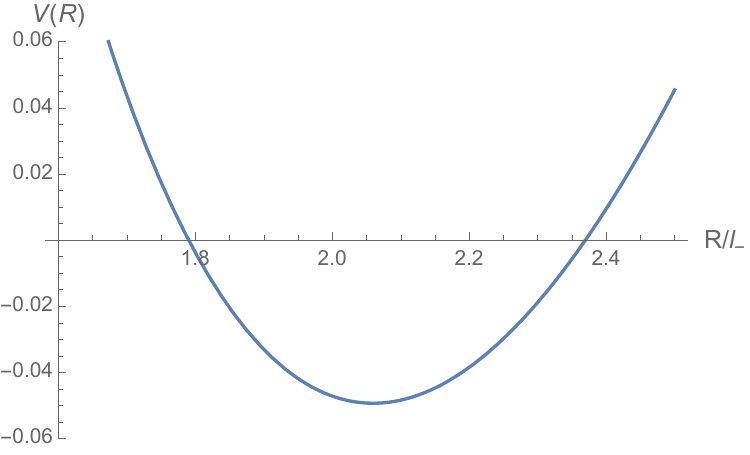}
  \end{center}
  \caption{
The potential form for $M_{+}=0.1$, $M_{-}=0.33$, $\tilde{a}_{+}=0$, $L_{+}=2.0$ and $s=0.2$
}
  \label{fig:potex}
\end{figure}

Let us calculate the
components of the symmetric tensor $K_{Eab}$.
From the $(\tau,\phi)$ component of Eq.~\eqref{eq:K}, we get 
\begin{align}
K_{E\tau\phi}
&=-4f_{E}J_{E}\frac{r}{R^2}\dot{t}^2_{E}-\frac{4J_{E}}{f_{E}}\frac{1}{r}\dot{R}^2, 
\label{eq:Ktp}
\end{align}
where 
we used the following expressions of each component of the Christoffel symbol
\begin{align}
\Gamma^{r}_{{t}_{E}\phi}&=4f_{E}J_{E}\frac{r}{R^2},  \\
\Gamma^{{t}_{E}}_{r\phi}&=-\frac{4J_{E}}{f_{E}}\frac{1}{r}.
\end{align}
Evaluating Eq.~\eqref{eq:Ktp} on the shell with Eq.~\eqref{eq:AB}, 
we get
\begin{align}
\left.K_{E\tau\phi}\right|_{\mathcal W}=-\frac{4J_{E}}{R}. 
\end{align}
Because $h_{E\tau\phi}=0$, the $(\tau,\phi)$ component of Eq.~\eqref{eq:2nd'} is equivalent to
\begin{align}
-\frac{4J_{E+}}{R}&+\frac{4J_{E-}}{R}=0, \nonumber \\
\therefore J_{+}&=J_{-}. 
\end{align}
Obviously, this is the consequence of the conservation of the angular momentum. 
The $(\phi,\phi)$ component of the extrinsic curvature is
\begin{align}
K_{E\phi\phi}
&=f_{E}r\dot{t}_{E}, 
\end{align}
where 
we used
\begin{align}
\Gamma^{r}_{\phi\phi}&=-f_{E}r.
\end{align}
By using Eq.~\eqref{eq:AB}, the 
$(\phi,\phi)$ component of Eq.~\eqref{eq:2nd'} can be written as
\begin{align}
\sqrt{f_{E+}-\dot{R}^2}-\sqrt{f_{E-}-\dot{R}^2}=-8\pi\sigma R. 
\end{align}
%where we used Eq.~(\ref{eq:AB}). 
Solving this for $\dot{R}^2$, we get
\begin{align}
-\frac{1}{2}\dot{R}^2&=V(R):=\frac{1}{2}\bar{\sigma}^2R^2-\frac{1}{2}\bar{f}_{E}+\frac{(\Delta f_{E})^2}{32\bar{\sigma}^2R^2},   
\end{align}
where, we defined $\bar{\sigma}:=4\pi\sigma$ and
\begin{align}
\bar{A}&:=\frac{A_{+}+A_{-}}{2}.
\end{align}
Therefore the shell dynamics reduces to the one-dimensional potential problem. 
The potential form can be rewritten as follows: 
\begin{align}
2\left(\frac{R}{l_-}\right)^2V(R)
&= A\left(\frac{R}{l_{-}}\right)^4+B\left(\frac{R}{l_{-}}\right)^{2}+C, \label{eq:V} 
\end{align}
where
\begin{align}
A&:=s^2-\frac{1}{2}\left(\frac{1}{L_{+}^2}+1\right)+\frac{1}{16s^2}\left(\frac{1}{L_{+}^2}-1\right)^2, \\
B&:=8\bar{M}-\frac{\Delta M}{s^2}\left(\frac{1}{L_{+}^2}-1\right), \\
C&:=-16\left(M_{+}L_{+}\tilde{a}_{+}\right)^2+\frac{4(\Delta M)^2}{s^2}. 
\end{align}
Here, we wrote $V(R)$ as a function of $R/l_{-}$ with the five dimensionless parameters: $M_{+}$, $M_{-}$, $\tilde{a}_{+}:=J/(M_{+}l_{+})$, 
$L_{+}:=l_{+}/l_{-}$ and $s:=\bar{\sigma}l_{-}$.
The functional form of $V(R)$ is depicted in Fig.~\ref{fig:potex} for $M_{+}=0.1$, $M_{-}=0.33$, $\tilde{a}_{+}=0$, $L_{+}=2.0$ and $s=0.2$. 
The Euclidean shell can oscillate between 
the two roots of $V(R)=0$, at which the velocity $\dot R$ of the shell vanishes. 
The 
bubble emerges with the radius determined by 
the condition $\dot{R}=0$ or equivalently $V(R)=0$. 
When we have two roots of $V(R)=0$, 
there are two possible radii of the bubble at the moment of the nucleation.
They share the bounce solution and have the same value of the nucleation rate. 
After the nucleation, the shell motion
obeys the junction condition in the Lorentz spacetime, and is described by the one-dimensional motion under the potential $-V(R)$. 
The shell is allowed to move in the region $-V(R)<0$.
The shell which emerges at the larger root of $V(R)=0$ expands, and the true vacuum region also expands. 
We call it an expanding solution. 
On the other hand, the shell which emerges at the smaller root shrinks, and the false vacuum region covers the whole region again. 
We call it a shrinking solution. 
It is worthy to note that, in the shrinking solution, 
through the shell accretion, the horizon area must increase due to the horizon area law, and 
the mass of the BH must be identical to $M_+$ due to the energy conservation. 
This fact indicates that the horizon area must decrease through the bubble nucleation. 
We will see this is actually realized in the current situation, and the decreasing horizon area 
indicates the quantum nature of the bubble nucleation (see App.~\ref{secC} for details.). 

\section{Conditions for the Existence of the Bounce Solution}
\label{sec3}

In order for the bubble nucleation process to exist and describe a physically reasonable process, 
several conditions must be satisfied. 
We list all the conditions and summarize the allowed region of the parameters in this section (see Refs.~\cite{Bachas:2021fqo,Bachas:2021tnp} for the coexistence of two BTZ spacetimes connected by thin-branes).

\subsection{Conditions for the Existence of Roots of $V(R)=0$}
\label{sec3.1}

For the shell to oscillate, 
we need a finite $V(R)<0$ region 
in $R>0$.
From Eq.~\eqref{eq:V}, we can regard $R^2V(R)$ as a quadratic function for $R^2$. 
For the functional form of $V(R)$ to be convex downward, the 
condition $A>0$ 
is required. 
In addition, 
we need two real positive roots of $R^2V(R)=0$ for $R^2$~(see App.~\ref{secD} for other possibilities). 
This condition requires $C>0$, $-B/2A>0$ and $B^2-4AC>0$. 
To summarize, we need 
$A\cdots D$ satisfying the following four conditions: 
\begin{align}
A>0,  \label{eq:A} \\
B<0,  \label{eq:B} \\
C>0,  \label{eq:C} \\
D:= B^2-4AC\geq0. \label{eq:D}
\end{align}

From Eq.~\eqref{eq:A}, we get
\begin{align} 
s^2-\frac{1}{2}\left(\frac{1}{L_{+}^2}+1\right)+\frac{1}{16s^2}\left(\frac{1}{L_{+}^2}-1\right)^2>0, 
\end{align}
or equivalently, 
\begin{align} 
s^2<\frac{1}{4L_{+}^2}-\frac{1}{2L_{+}}+\frac{1}{4},\frac{1}{4L_{+}^2}+\frac{1}{2L_{+}}+\frac{1}{4}<s^2. \label{eq:A'}
\end{align}
Eq.~\eqref{eq:B} requires $(\Delta \Lambda)(\Delta M)<0$. 
In vacuum decay, we assume $\Lambda_{+}>\Lambda_{-}$, and obtain $M_{+}<M_{-}$ as a necessary condition. 
In other words, the mass must 
decrease 
to form a bounce solution. 
Moreover, Eq.~\eqref{eq:B} 
imposes the lower bound for $M_{-}$:
\begin{align}
M_{-}>M_{B}:=M_{+}\frac{1-L_{+}^2(1+4s^2)}{1-L_{+}^2(1-4s^2)} .\label{eq:B'}
\end{align}
Eq.~\eqref{eq:C} can be written as follows: 
\begin{align} 
-16\left(M_{+}L_{+}\tilde{a}_{+}\right)^2+\frac{4(\Delta M)^2}{s^2}>0.  
\end{align}
This gives 
the condition for $M_{-}$ under fixed values of $L_{+}$, $s$, $\tilde{a}_{+}$ and $M_{+}$:
\begin{align}
M_{-}>M_{C}:=M_{+}(1+2sL_{+}\tilde{a}_{+}) . \label{eq:C'}
\end{align}
Another condition for the mass also stems from Eq.~(\ref{eq:D}): 
\begin{align}
M_{-}\geq
&M_{D}:=\frac{M_{+}}{2}\left[1+L_{+}^2(1-4s^2)+\sqrt{(1-\tilde{a}_{+}^2)\{1+L_{+}^4(1-4s^2)^2-2L_{+}^2(1+4s^2)\}}\right] ,\nonumber\\
M_{-}\leq
&M'_{D}:=\frac{M_{+}}{2}\left[1+L_{+}^2(1-4s^2)-\sqrt{(1-\tilde{a}_{+}^2)\{1+L_{+}^4(1-4s^2)^2-2L_{+}^2(1+4s^2)\}}\right] . \label{eq:D'}
\end{align}
When $D=0$, the shell can exist only at the radius of the degenerate root of $V(R)=0$, that is, 
the shell has a constant radius with time. We call it a static shell.

\subsection{Constraints from Time Parametrization}
\label{sec3.2}

In addition to
the conditions 
for the potential form,
there 
are
other conditions
for the motion of the shell 
to be physical. 
Specifically, the shell motion must be ``future-directed" in both sides of the shell, 
that is, $\dot{t}_{E\pm}\geq0$.
We can rewrite these requirements as
\begin{align}
f_{E+}(R)\dot{t}_{E+}&=\sqrt{f_{E+}(R)-\dot{R}^2}=-\bar{\sigma}R-\frac{\Delta f_{E}}{4\bar{\sigma}R} \geq0,
\label{eq:outer}  \\
f_{E-}(R)\dot{t}_{E-}&=\sqrt{f_{E-}(R)-\dot{R}^2}=\bar{\sigma}R-\frac{\Delta f_{E}}{4\bar{\sigma}R} 
\label{eq:inner} \geq0,
\end{align}
where we used Eq.~\eqref{eq:AB}.

The inequality \eqref{eq:outer} is equivalent to
\begin{align}
\left(\frac{R}{l_{-}}\right)\left[-s-\frac{L_{+}^{-2}-1}{4s}\right]+\frac{1}{R/l_{-}}\frac{2\Delta M}{s}\geq0.
\label{eq:ineq1}
\end{align}
From the discussion in the previous subsection, 
we necessarily have $\Delta M<0$. 
Hence, for the inequality \eqref{eq:ineq1} to be satisfied, the coefficient of $R$ must be positive. 
From this, we obtain the constraint for $s$:
\begin{align}
s^2<\frac{1}{4}-\frac{1}{4L_{+}^2} .
\label{eq:s'^2}
\end{align}

The inequality \eqref{eq:ineq1} is
saturated when
\begin{align}
\frac{R}{l_{-}}=\frac{R^*_{+}}{l_{-}}:=\sqrt{\frac{2\Delta M}{s\left(s+\frac{L_{+}^{-2}-1}{4s}\right)}},
\end{align}
and 
the shell must oscillate between 
two radii
larger than $R^*_{+}$. This requires $(R^*_{+})^2V(R^*_{+})>0$ and $\left.\frac{d}{dR}\left[R^2V(R)\right]\right|_{R=R^*_{+}}<0$. 
These conditions are equivalent to
\begin{align}
&\frac{\tilde{a}_{+}^2M_{+}^2 (-1+L_{+}^2(1-4s^2))^2+4(M_{-}-M_{+})(M_{-}+L_{+}^2M_{+}(-1+4s^2))}{(M_{-}-M_{+})  
(1+L_{+}^2(-1+4s^2))}>0, \label{eq:dt1} \\
&4M_{+}+\frac{8(M_{-}-M_{+})}{1+L_{+}^2(-1+4s^2)}<0. \label{eq:dt2} 
\end{align}

We may consider the condition \eqref{eq:inner} in a similar way to that for the condition \eqref{eq:outer}. 
First, we rewrite the inequality \eqref{eq:inner} as
\begin{align}
\left(\frac{R}{l_{-}}\right)\left[s-\frac{L_{+}^{-2}-1}{4s}\right]+\frac{1}{R/l_{-}}\frac{2\Delta M}{s}\geq0 . \label{eq:ineq2}
\end{align}
We find that the coefficient of $R$ is positive
for any $s>0$. 
The inequality
\eqref{eq:ineq2} is saturated when
\begin{align}
\frac{R}{l_{-}}=\frac{R^*_{-}}{l_{-}}:=\sqrt{\frac{-2\Delta M}{s\left(s-\frac{L_{+}^{-2}-1}{4s}\right)}}, 
\end{align}
and the region of the motion must be larger than $R^*_{-}$. 
Hence we require $(R^*_{-})^2V(R^*_{-})>0$ and $\frac{d}{dR}\left[R^2V(R)\right]|_{R=R^*_{-}}<0$. 
However, we
see that $\frac{R^*_{+}}{l_{-}}>\frac{R^*_{-}}{l_{-}}$ from
\begin{align}
\left(\frac{R^*_{+}}{l_{-}}\right)^2-\left(\frac{R^*_{-}}{l_{-}}\right)^2
=\frac{4\Delta M}{\left(s+\frac{L_{+}^{-2}-1}{4s}\right)\left(s-\frac{L_{+}^{-2}-1}{4s}\right)}>0.
\end{align}
Therefore
it is sufficient to require only $(R^*_{+})^2V(R^*_{+})>0$ and $\frac{d}{dR}\left[R^2V(R)\right]|_{R=R^*_{+}}<0$. 
By solving 
Eqs.~(\ref{eq:dt1}) and (\ref{eq:dt2}), 
we obtain
\begin{align}
&M_{-}<M_{T}:=\frac{M_{+}}{2}\left[1+L_{+}^2(1-4s^2)+\sqrt{(1-\tilde{a}_{+}^2)\{1+L_{+}^2(-1+4s^2)\}^2}\right], \label{eq:VRS} \\
&M_{-}>M'_{T}:=\frac{M_{+}}{2}\left[1+L_{+}^2(1-4s^2)\right]. \label{eq:DVRS}
\end{align}

\subsection{Constraints from Horizon Condition}
\label{sec3.3}

In addition to the above conditions, we 
impose the condition that
the shell 
oscillates outside the event horizon. 
The outer horizon radius 
is given by
the roots of $f_{E}(r)=0$, which can be written as%
\footnote{
Subscript $+$/$-$ 
indicates the horizon radius before/after the nucleation, not the outer/inner horizon. 
From this expression, we can see that the condition for the extremal BH is $J/l=M$.}
\begin{align}
r_{\mathcal{H}_{\pm}}=\sqrt{4M_{\pm}l_{\pm}^2+4l_{\pm}\sqrt{M_{\pm}^2l_{\pm}^2-J^2}} .
\end{align}
Since 
$r_{\mathcal{H}_{-}}<r_{\mathcal{H}_{+}}$ from 
other conditions~(see App.~\ref{secC}), it is sufficient to 
require that the shell 
oscillates between two radii larger than $r_{\mathcal{H}_{+}}$. 
That is, we impose the following conditions: 
$V(r_{\mathcal{H}_{+}})\geq0$ and $(r_{\mathcal{H}_{+}})^2<-B/2A$. 
While the former puts no constraint on the parameters, from the latter, we obtain the inequality for $M_{-}$:
\begin{align}
M_{-}>M_{R}:=-M_{+}\frac{-1+\sqrt{1-\tilde{a}_{+}^2}+(1+\sqrt{1-\tilde{a}_{+}^2})L_{+}^4(1-4s^2)^2-2\sqrt{1-\tilde{a}_{+}^2}L_{+}^2(1+4s^2)}{2+L_{+}^2(-2+8s^2)}.
\label{eq:MR2}
\end{align}

\subsection{Allowed Region in the Parameter Space}
\label{sec3.4}

In this subsection, 
we summarize the above conditions and show some examples of the allowed region
for the parameters. 
First, we 
consider the conditions~\eqref{eq:A'} and \eqref{eq:s'^2} for $L_{+}$ and $s$, 
and fix them. 
Next, we fix $\tilde{a}_{+}$, and show the allowed region for $M_{+}$ and $M_{-}$ %from 
by considering
Eqs.~\eqref{eq:B'}, \eqref{eq:C'}, \eqref{eq:D'} and \eqref{eq:VRS}.

\begin{figure}[H]
    \begin{tabular}{cc}
      \begin{minipage}[t]{0.45\hsize}
        \centering
  \includegraphics[clip,width=8cm]{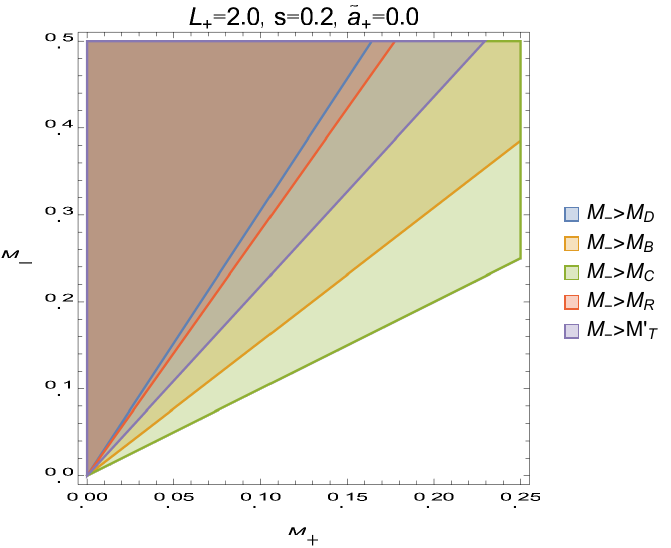}
  \caption{The lower bound of $M_{-}$}
  \label{fig:ssJ0}
      \end{minipage} &
      \begin{minipage}[t]{0.45\hsize}
  \includegraphics[clip,width=8cm]{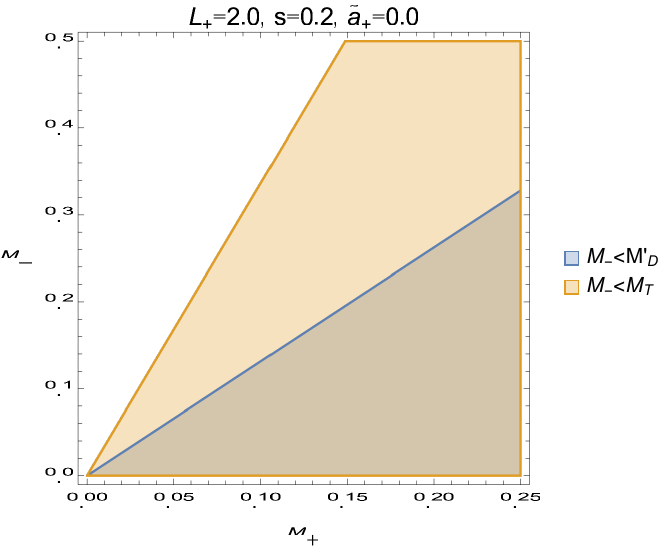}
  \caption{The upper bound of $M_{-}$}
  \label{fig:dottJ0}
      \end{minipage}
    \end{tabular}
  \end{figure}

On
the two conditions of~(\ref{eq:A'}), 
only the second one,
$s^2<\frac{1}{4L_{+}^2}-\frac{1}{2L_{+}}+\frac{1}{4}$, 
can be
compatible with the condition (\ref{eq:s'^2}). 
With $L_{+}>1$, we
see that
\begin{align}
\left(-\frac{1}{4L_{+}^2}+\frac{1}{4}\right)-\left(\frac{1}{4L_{+}^2}-\frac{1}{2L_{+}}+\frac{1}{4}\right)=\frac{1}{2}\left(\frac{1}{L_{+}}-\frac{1}{L^2_{+}}\right)>0 . \label{eq:sineq}
\end{align}
By using this Eq.~(\ref{eq:sineq}) and $s>0$, we eventually find 
\begin{align}
s<\frac{1}{2}-\frac{1}{2L_{+}} \label{eq:sfin}
\end{align}
for a given value of 
$L_{+}$. 
From this condition, we can simplify the expression $M_T$ as 
\begin{align}
M_{T}=\frac{M_{+}}{2}\left[1+L_{+}^2(1-4s^2)-\sqrt{1-\tilde{a}_{+}^2}\left\{1+L_{+}^2(-1+4s^2)\right\}\right].
\label{eq:Mt}
\end{align}
Setting $L_{+}=2.0$, $s=0.2$ and $\tilde{a}_{+}=0.0$, 
we obtain 
the allowed regions indicated by 
the conditions~(\ref{eq:B'}), (\ref{eq:C'}), (\ref{eq:D'}) and (\ref{eq:VRS}) 
as
the shaded regions in Figs.~\ref{fig:ssJ0} and Fig.\ref{fig:dottJ0}.
From Figs.~\ref{fig:ssJ0} and \ref{fig:dottJ0}, we see that the allowed region cannot 
exist in the region $M_{-}<M'_{D}$ and 
we have to take the branch $M_{-}>M_{D}$. 

\begin{figure}[H]
\begin{center}
  \includegraphics[clip,width=6cm]{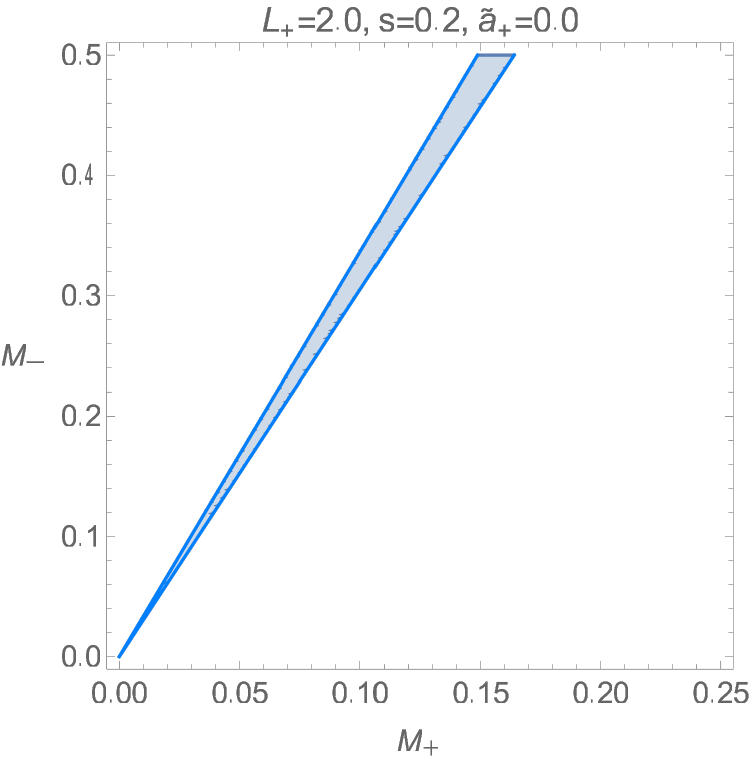}
  \end{center}
  \caption{The allowed region ($\tilde{a}_{+}=0.0$)}
  \label{fig:condJ0}
\end{figure}

\begin{figure}[H]
    \begin{tabular}{cc}
      %---- 最初の図 ---------------------------
      \begin{minipage}[t]{0.45\hsize}
        \centering
  \includegraphics[clip,width=6cm]{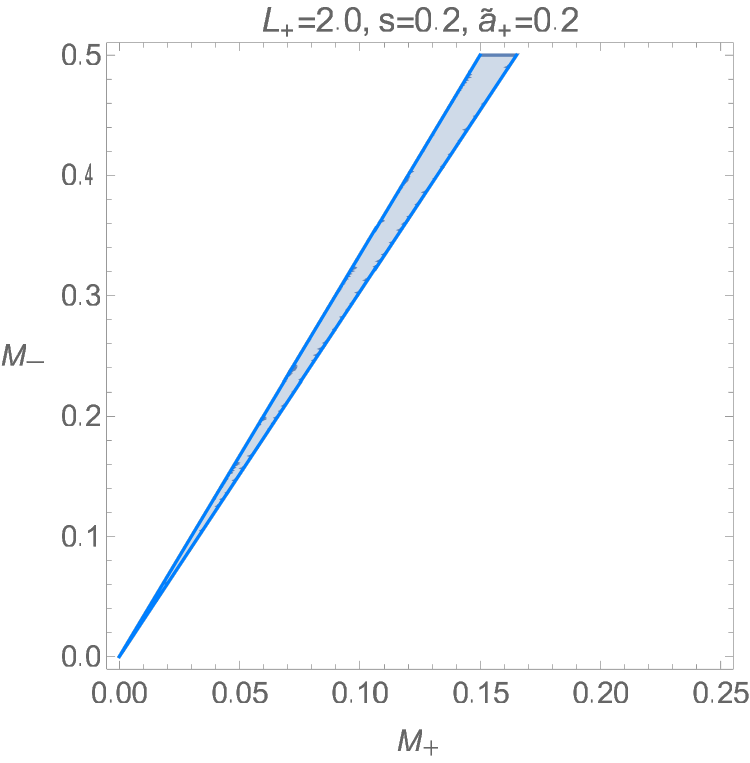}
      \end{minipage} &
      %---- 2番目の図 --------------------------
      \begin{minipage}[t]{0.45\hsize}
  \includegraphics[clip,width=6cm]{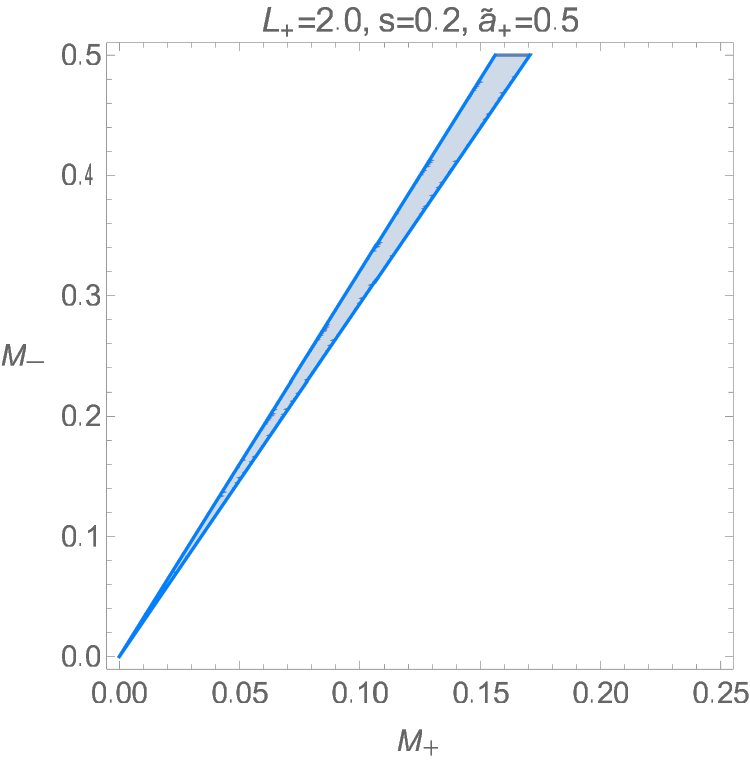}
      \end{minipage}
      %---- 図はここまで ----------------------
    \end{tabular}
     \caption{The allowed region ($\tilde{a}_{+}\neq0.0$)} \label{fig:ssa}
  \end{figure}

By combining them we eventually obtain Fig.~\ref{fig:condJ0} as the allowed region for the masses. 
With a non-zero angular momentum, 
for example 
$\tilde{a}_{+}=0.2$, $0.5$,
we get 
Fig.~\ref{fig:ssa}.
Regardless of the value of the angular momentum, 
the lower bound of $M_-$ is given by that of
the static shell.

\section{Euclidean Action and the Decay Rate} 
\label{sec4}

According to Ref.~\cite{Coleman:1977py}, 
the decay rate per unit of time and volume $\Gamma$ is given by
\begin{align}
\Gamma\propto e^{-\mathcal{B}/\hbar} ,
\end{align}
where the value of $\mathcal{B}$ 
is given by the difference between the values of the Euclidean action of the 
bounce solution $S_{E}$ and the false vacuum $S_{E0}$ as $\mathcal{B}:= S_{E}-S_{E0}$.
Since the main contribution to the decay rate comes from the exponential part, 
ignoring the factor of the dependence, we focus on the value of $\mathcal B$.
Then we can estimate the decay rate by calculating the Euclidean actions. 
In this section, we derive an analytic form of $\mathcal{B}$ in the BTZ case.

$S_{E}$ can be divided into the contributions from the horizon $S_{\mathcal H}$, 
bubble wall $S_{\mathcal W}$ and bulk $S_{\mathcal M_\pm}$ as follows:
\begin{align}
S_{E}=S_{\mathcal{H}}+S_{\mathcal{M}_{+}}+S_{\mathcal{M}_{-}}+S_{\mathcal{W}}.
\end{align}
$S_{\mathcal{H}}$ can be evaluated by considering the action in the vicinity of the horizon (see Appendices in Refs.~\cite{Gregory:2013hja,Oshita:2019jan}). 
In the current setting, the value of $S_{\mathcal{H}}$ is given as  
\begin{align}
S_{\mathcal{H}}&=-\frac{A_{\mathcal{H}_{-}}}{4}, 
\end{align}
where 
$A_{\mathcal{H}_{-}}$ is 
the area of the horizon. 
$S_{\mathcal{W}}$ 
can be written as
\begin{align}
S_{\mathcal{W}}&=-\int_{\mathcal{W}}d^2x\sqrt{h_{E}}\int_{R-0}^{R+0}dr{\mathcal{L}_{m}} \nonumber \\
&
=\int_{\mathcal{W}}d^2x\sqrt{h_{E}}\int_{R-0}^{R+0}dr\sigma\delta(r-R(t_E)) \nonumber \\
&=\int_{\mathcal{W}}d^2x\sqrt{h_{E}}\sigma. \label{eq:Sw}
\end{align}
$S_{\mathcal{M}_{\pm}}$ 
can be written as 
\begin{align}
S_{\mathcal{M}_{\pm}}&=-\int_{\mathcal{M}_{\pm}}d^3x\sqrt{g_{E\pm}}\left(\frac{1}{16\pi}^3\mathcal{R}^{(E)}+\mathcal{L}_{m}^{(E)}\right)+\frac{1}{8\pi}\int_{\partial\mathcal{M}_{\pm}}d^2x\sqrt{h_{E}}\tilde{K}_{E\pm}, 
\end{align}
where the second term is the boundary term. 
With the ADM decomposition, we obtain 
\begin{align}
S_{\mathcal{M}_{\pm}}&=-\frac{1}{16\pi}\oint dt_{E\pm}\int_{\Sigma_{t_{E\pm}}}d^2x\sqrt{g_{E\pm}}\left(^2\mathcal{R}^{(E)}-\tilde{K}^2_{E\pm}+\tilde{K}_{E\pm ab}\tilde{K}_{E\pm}^{ab}+16\pi\mathcal{L}_{m}^{(E)}\right)  \nonumber \\
&+\frac{1}{8\pi}\int_{\mathcal{W}}d^2x\sqrt{h_{E}}\tilde{K}_{E\pm}+\frac{1}{8\pi}\int_{\mathcal{W}}d^2x\sqrt{h_{E}}\tilde{n}_{\pm\mu}\tilde{u}^{\nu}_{\pm}\nabla_{\nu}\tilde{u}^{\mu}_{\pm}, 
\end{align}
where $\tilde{u}^{\mu}_{\pm}$ is the unit normal vector of the constant time slice 
$\Sigma_{t_{E\pm}}$, $\tilde{n}_{\pm\mu}$ is the ingoing  unit normal vector of 
$\mathcal{W}$, 
and 
$\tilde{K}_{E\pm}=\pm K_{\pm}$ is the Euclidean extrinsic curvature 
associated with $\tilde{n}_{\pm\mu}$. The first line vanishes from the Hamiltonian constraint 
$ ^2\mathcal{R}^{(E)}-\tilde{K}^2_{E\pm}+\tilde{K}_{E\pm ab}\tilde{K}_{E\pm}^{ab}+16\pi\mathcal{L}_{m}^{(E)}=0$, 
and we obtain
\begin{align}
S_{\mathcal{M}_{+}}+S_{\mathcal{M}_{-}}&=\frac{1}{8\pi}\int_{\mathcal{W}}d^2x\sqrt{h_{E}}(K_{+}-K_{-})+\frac{1}{8\pi}\int_{\mathcal{W}}d^2x\sqrt{h_{E}}(n_{+\mu}\tilde{u}^{\nu}_{+}\nabla_{\nu}\tilde{u}^{\mu}_{+}-n_{-\mu}\tilde{u}^{\nu}_{-}\nabla_{\nu}\tilde{u}^{\mu}_{-}) .\label{eq:SM}
\end{align}

In the BTZ case, the unit normals are given by 
\begin{align}
\tilde{u}^{\mu}_{\pm}&=\left(\sqrt{g_{E\pm}^{tt}},0,\frac{g_{E\pm}^{t\phi}}{\sqrt{g_{E\pm}^{tt}}}\right), \\
\tilde{n}_{\pm\mu}&=\left(\mp\dot{R},\pm\dot{t}_{E},0\right). 
\end{align}
Then, on $\mathcal W$, $\tilde u^\phi_\pm=0$, and we obtain 
\begin{align}
\left.n_{\pm\mu}\tilde{u}^{\nu}_{\pm}\nabla_{\nu}\tilde{u}^{\mu}_{\pm}\right|_{\mathcal W}&=\left.\left(
-\dot{R}\tilde{u}^{t_{E}}\nabla_{t_E}\tilde{u}^{t_{E}}+\dot{t}_{E}\tilde{u}^{t_{E}}\nabla_{t_E}\tilde{u}^{r}\right)\right|_{\mathcal W} \nonumber \\
&=\left.-\frac{\partial_{r}f_{E\pm}}{2}\dot{t}_{E\pm}\right|_{\mathcal W} , 
\end{align}
where
we used 
\begin{align}
\Gamma^{r}_{{t}_{E}{t}_{E}}&=-\frac{f_{E}}{2}\partial_{r}f_{E}.
\end{align}

By using the trace of the 2nd junction condition (\ref{eq:2nd})
\begin{align}
[K]_{\pm}=-16\pi\sigma,\label{eq:2ndT}
\end{align}
we get 
\begin{align}
S_{\mathcal{M}_{+}}+S_{\mathcal{M}_{-}}&=-2\int_{\mathcal{W}}d^2xR\sigma-\frac{1}{16\pi}\int_{\mathcal{W}}d^2xR(\partial_{r}f_{E+}\dot{t}_{E+}-\partial_{r}f_{E-}\dot{t}_{E-}).
\end{align}
Summing up all contributions, we rewrite the Euclidean action as
\begin{align}
S_{E}&=-\frac{A_{\mathcal{H}_{-}}}{4}-2\int_{\mathcal{W}}d^2xR\sigma-\frac{1}{16\pi}\int_{\mathcal{W}}d^2xR(\partial_{r}f_{E+}\dot{t}_{E+}-\partial_{r}f_{E-}\dot{t}_{E-})+\int_{\mathcal{W}}d^2xR\sigma \nonumber \\
&=-\frac{A_{\mathcal{H}_{-}}}{4}-\frac{1}{16\pi}\int_{\mathcal{W}}d^2x[(R\partial_{r}f_{E+}-2f_{E+})\dot{t}_{E+}-(R\partial_{r}f_{E-}-2f_{E-})\dot{t}_{E-}] \nonumber \\
&=-\frac{A_{\mathcal{H}_{-}}}{4}-\frac{1}{8}\oint d\tau[(R\partial_{r}f_{E+}-2f_{E+})\dot{t}_{E+}-(R\partial_{r}f_{E-}-2f_{E-})\dot{t}_{E-}]  \nonumber \\
&=-\frac{A_{\mathcal{H}_{-}}}{4}-2\oint d\tau\left[\left(M_{+}+\frac{4J^2_{E+}}{R^2}\right)\dot{t}_{E+}-\left(M_{-}+\frac{4J^2_{E-}}{R^2}\right)\dot{t}_{E-}\right] ,\label{eq:cor}
\end{align}
where
we used the relation
\begin{align}
R\sigma=-\frac{1}{8\pi}[f_{E}\dot{t}_{E}]_{\pm}
\end{align}
given by the $(\phi,\phi)$ component of Eq.~\eqref{eq:2nd'}. 
By using $S_{E0}=-\frac{A_{\mathcal{H}_{+}}}{4}$, we obtain
\begin{align}
\mathcal{B}=\frac{A_{\mathcal{H}_{+}}}{4}-\frac{A_{\mathcal{H}_{-}}}{4}-2\oint d\tau\left[\left(M_{+}-\frac{4J^2}{R^2}\right)\dot{t}_{E+}-\left(M_{-}-\frac{4J^2}{R^2}\right)\dot{t}_{E-}\right] .
\end{align}

The horizon area is
\begin{align}
A_{\mathcal{H}_{\pm}}&=\int_{0}^{2\pi}d\phi r_{\mathcal{H}_{\pm}} \nonumber \\
&=2\pi r_{\mathcal{H}_{\pm}}.
\end{align}
Then we can rewrite $\mathcal{B}$ as
\begin{align}
\mathcal{B}&=\frac{\pi}{2}(r_{\mathcal{H}_{+}}-r_{\mathcal{H}_{-}})-2\oint d\tau\left[\left(M_{+}-\frac{4J^2}{R^2}\right)\dot{t}_{E+}-\left(M_{-}-\frac{4J^2}{R^2}\right)\dot{t}_{E-}\right] .\label{eq:BI}
\end{align}

\section{Results} 
\label{sec5}

In this section, we show the parameter dependence of the nucleation rate for the cases 
which satisfy the conditions listed in Sec.~\ref{sec3}. 
First, we evaluate 
the nucleation rate for 
$\tilde{a}_{+}=0$ 
and clarify the case which 
gives the dominant contribution to the tunnelling. 
Next, we introduce the angular momentum and 
see how the nucleation rate of the dominant case changes depending on the value of 
$\tilde{a}_{+}$. 

\subsection{$\tilde{a}_{+}=0$ case}
\label{sec5.1}

\begin{figure}[H]
\begin{center}
  \includegraphics[clip,width=10cm]{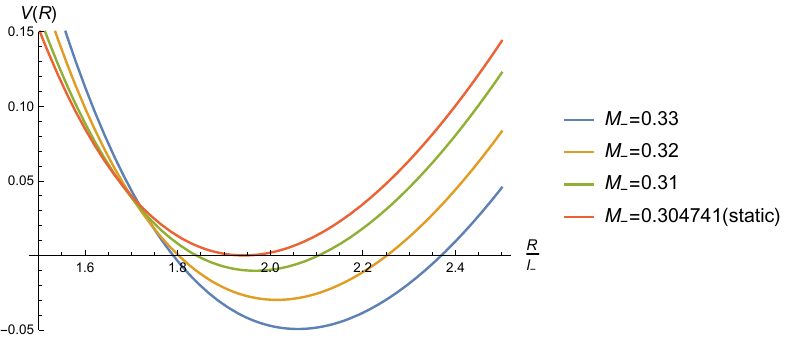}
  \end{center}
  \caption{The potential form for $\tilde{a}_{+}=0,M_{+}=0.1,L_{+}=2.0,s=0.2$}
  \label{fig:J=0}
\end{figure}

First, we set $\tilde{a}_{+}=0$. 
Setting $M_{+}=0.1$, $L_{+}=2.0$ and $s=0.2$, 
we plot the potential form for each value of $M_-$ in Fig.~\ref{fig:J=0}.
The available region for the shell motion gets smaller for a smaller value of $M_-$. 

\begin{figure}[H]
 \begin{minipage}[t]{0.45\hsize}
  \centering
   \includegraphics[clip,width=7cm]{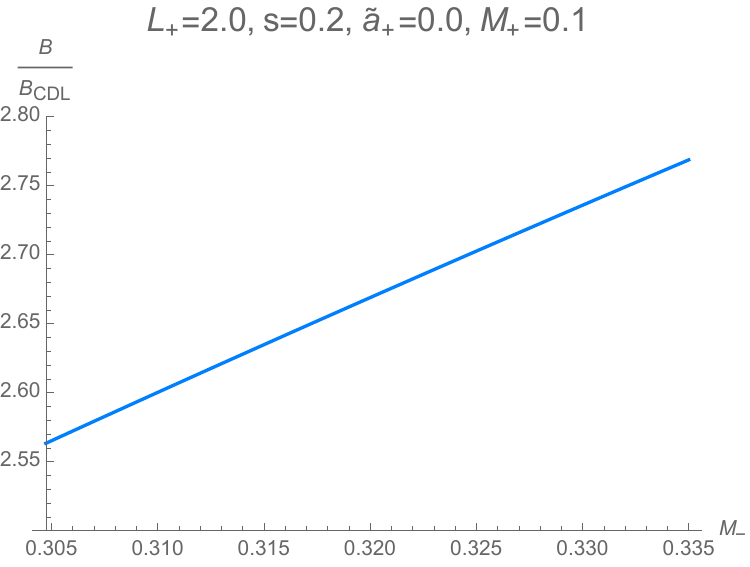}
  \subcaption{}\label{fig:J0l2M01}
 \end{minipage}
 \begin{minipage}[t]{0.45\hsize}
  \centering
  \includegraphics[clip,width=7cm]{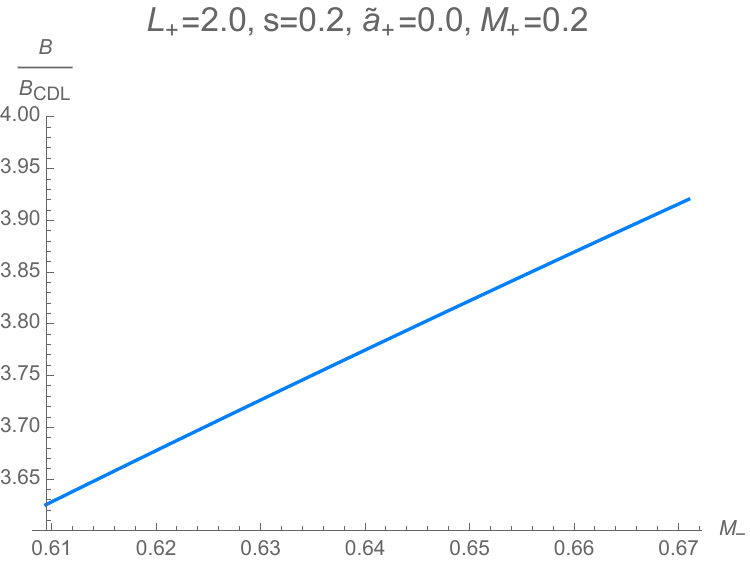}
  \subcaption{}\label{fig:J0l2M02}
 \end{minipage}\\
 \begin{minipage}[t]{0.45\hsize}
  \centering
  \includegraphics[clip,width=7cm]{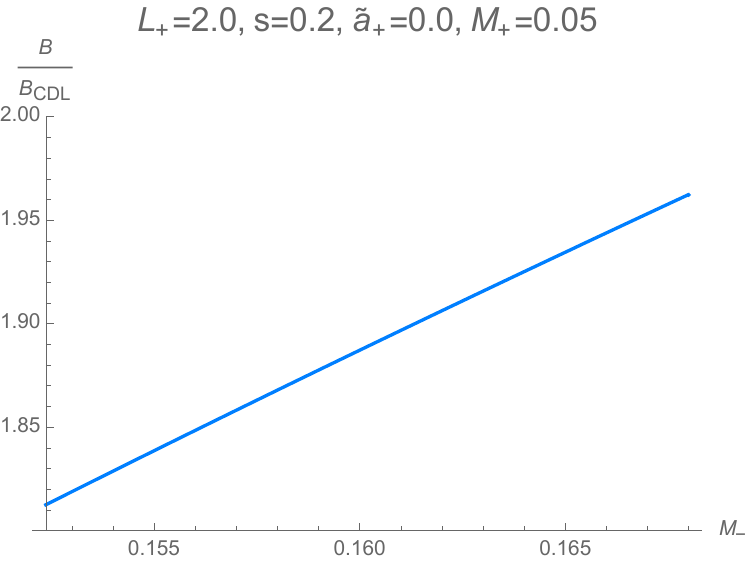}
  \subcaption{}\label{fig:J0l2M005}
 \end{minipage}
 \begin{minipage}[t]{0.45\hsize}
  \centering
  \includegraphics[clip,width=7cm]{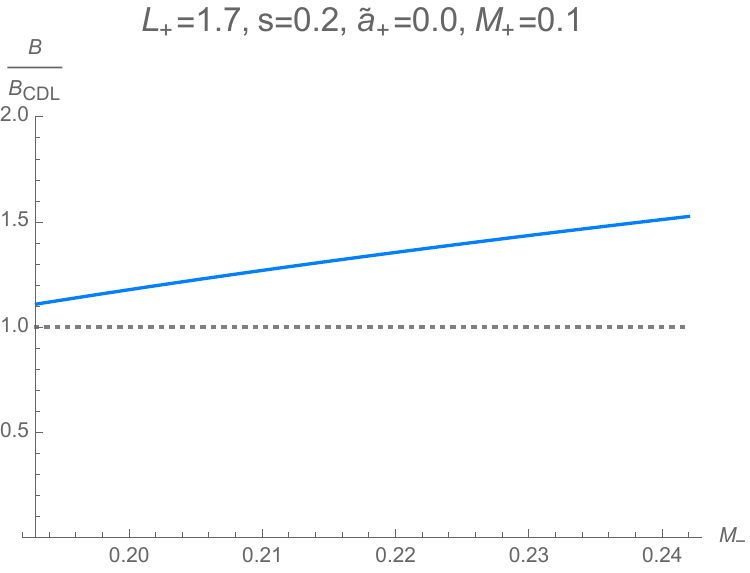}
  \subcaption{}\label{fig:J0l17}
 \end{minipage}
 \caption{The $M_{-}$ dependence of $\mathcal{B}$ ($\tilde{a}_{+}=0$)}\label{fig:B2}
\end{figure}

Setting $\tilde{a}_{+}=0$, $L_{+}=2.0$ and $s=0.2$, 
we show the $M_-$ dependence of $\mathcal{B}/\mathcal{B}_{CDL}$ for $M_+=0.1$, $0.2$ and $0.05$ 
in Figs.~\ref{fig:J0l2M01}, \ref{fig:J0l2M02} and \ref{fig:J0l2M005}, respectively (see App.~\ref{secA} for the definition of $\mathcal{B}_{CDL}$). 
For all the cases, the left boundary of the plot region is set to the value of $M_-$ for the static shell. 
We can see that, with a fixed value of $M_{+}$, the static shell case has the largest nucleation probability, 
and 
the rate is larger for a smaller value of $M_{+}$. 
This result does not change even if we change the value of the AdS length $L_+$ as is explicitly shown for the case $M_{+}=0.1$, $L_{+}=1.7$ and $s=0.2$ 
in Fig.~\ref{fig:J0l17}. 
The $L_{+}=1.7$ case has a
smaller value of 
$\mathcal{B}/\mathcal{B}_{CDL}$ than the $L_{+}=2.0$ case. 
We note that the value of $\mathcal{B}/\mathcal{B}_{CDL}$ is 
larger than unity in all cases shown in Figs.~\ref{fig:J0l2M01}--\ref{fig:J0l17}.
%This result 
%indicates that the existence of a BH can decrease the decay rate, 
%as opposed to the results 
%reported in Refs.~\cite{Gregory:2013hja,Burda:2015isa}, which state that BHs act as nucleation sites. 

\begin{figure}[H]
 \begin{minipage}[t]{0.45\hsize}
  \centering
   \includegraphics[clip,width=7.5cm]{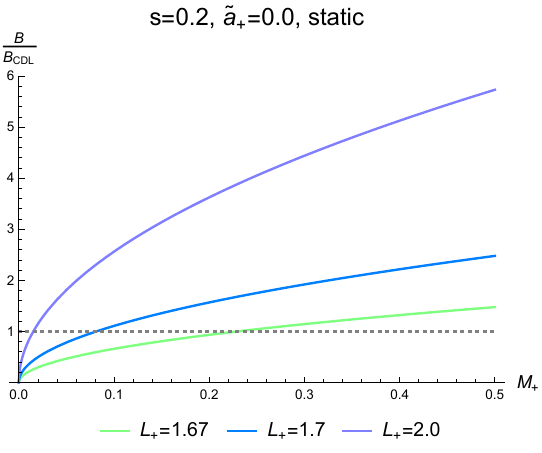}
  \subcaption{}\label{fig:J0s02}
 \end{minipage}
 \begin{minipage}[t]{0.45\hsize}
  \centering
  \includegraphics[clip,width=7.5cm]{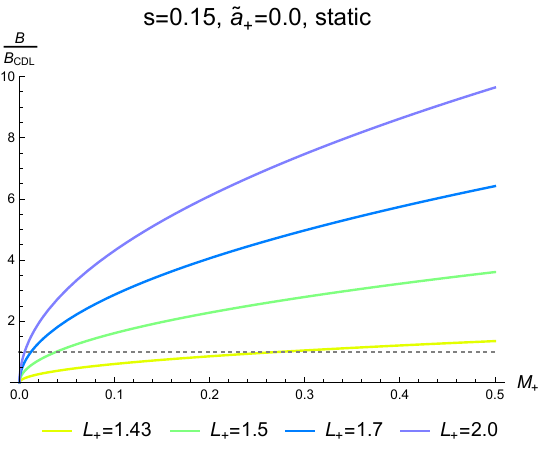}
  \subcaption{}\label{fig:J0s15}
 \end{minipage}\\
 \caption{The $M_{+}$ dependence of $\mathcal{B}$ of the static shells}\label{fig:stJ=0}
\end{figure}

The $M_{+}$ dependence of $\mathcal{B}$ in the case of the static shell 
for each value of $L_+$ is
shown in Figs.~\ref{fig:J0s02} and \ref{fig:J0s15}
with $s=0.2$ and $s=0.15$, respectively. 
Here, we note that the parameter sets of 
$(s,L_{+})=(0.2, 1.67)$ and $(0.15, 1.43)$ 
almost saturate the inequality (\ref{eq:sfin}).
We can see that, with fixed $s$ and $L_{+}$, $\mathcal{B}$ in the static shell case increases monotonically with $M_{+}$.
When the mass of the seed BH is small enough $\left(M_+\lesssim 0.02\right)$,
we 
find $\mathcal{B}<\mathcal{B}_{CDL}$. 
This tendency is similar to the results in Refs.~\cite{Gregory:2013hja,Burda:2015isa}. 
We can see that when we take the same values of $L_{+}$ and $M_{+}$,  
a smaller value of $s$ gives a larger value of $\mathcal{B}/\mathcal{B}_{CDL}$. 
It should be noted that, in our case, differently from Refs.~\cite{Gregory:2013hja,Burda:2015isa}, 
the mass necessary increase through the nucleation, and the static shell case always exists. 

\subsection{$\tilde{a}_{+}\neq0$ cases}
\label{sec5.2}

\begin{figure}[H]
\begin{center}
  \includegraphics[clip,width=9cm]{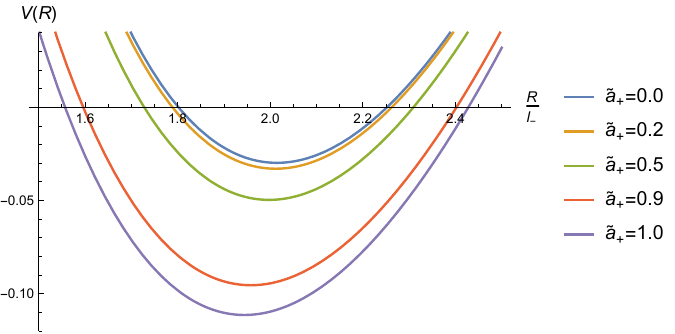}
  \end{center}
  \caption{The potential form for $M_{+}=0.1$, $M_{-}=0.32$, $L_{+}=2.0$ and $s=0.2$ } 
%($\tilde{a}_{+}\neq0$)}
  \label{fig:Jns}
\end{figure}

\begin{figure}[H]
    \begin{tabular}{cc}
      %---- 最初の図 ---------------------------
      \begin{minipage}[t]{0.45\hsize}
        \centering
        \includegraphics[clip,width=7cm]{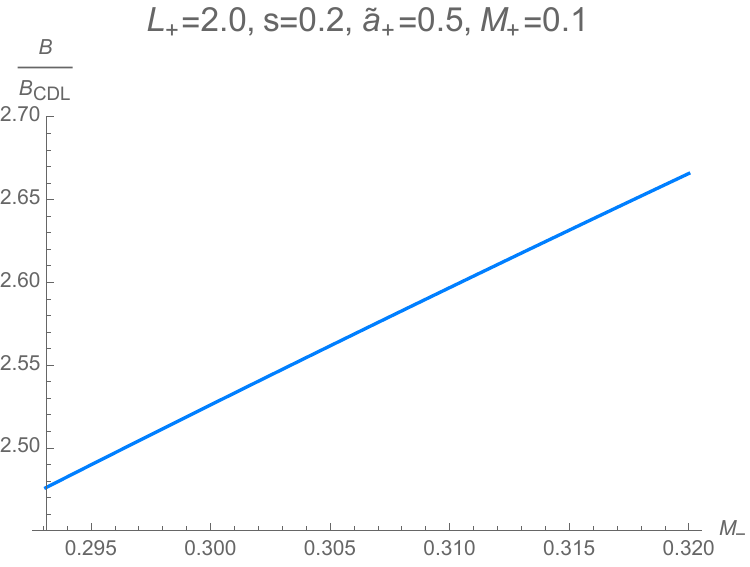}
      \end{minipage} &
      %---- 2番目の図 --------------------------
      \begin{minipage}[t]{0.45\hsize}
         \includegraphics[clip,width=7cm]{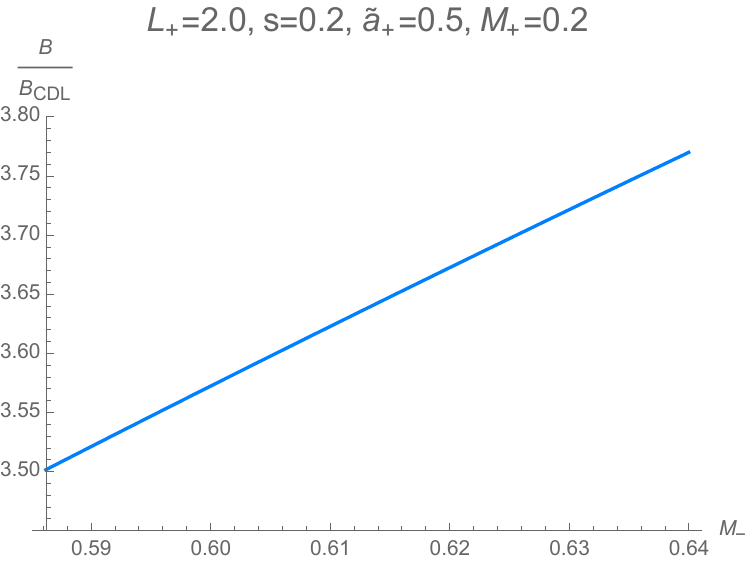}
      \end{minipage}
      %---- 図はここまで ----------------------
    \end{tabular}
    \caption{The $M_{-}$ dependence of $\mathcal{B}$ ($\tilde{a}_{+}\neq0$)} \label{fig:Jnz}
  \end{figure}

Next, we consider $\tilde{a}_{+}\neq0$ cases. 
The potential form with $M_{+}=0.1$, $M_{-}=0.32$, $L_{+}=2.0$ and $s=0.2$ is shown for each value of 
$\tilde{a}_{+}$ in Fig.~\ref{fig:Jns}. 
We can see that increasing $\tilde{a}_{+}$ deepens the potential depth.
The numerical result for $\mathcal{B}$ is 
shown in Fig.~\ref{fig:Jnz}. 
The behavior is similar to the $\tilde{a}_{+}=0$ cases, 
and the formation of 
the static shell dominates the nucleation probability.

\begin{figure}[htbp]
\begin{center}
  \includegraphics[clip,width=8cm]{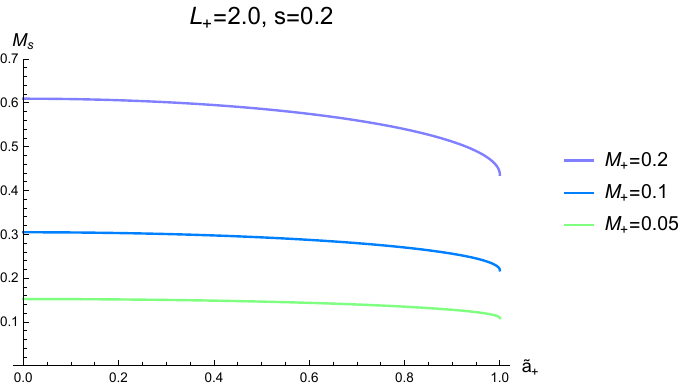}
  \end{center}
  \caption{The $\tilde{a}_{+}$ dependence of $M_{-}$ of static shells}
  \label{fig:stm}
\end{figure}

\begin{figure}[htbp]
    \begin{tabular}{cc}
      %---- 最初の図 ---------------------------
      \begin{minipage}[t]{0.45\hsize}
        \centering
        \includegraphics[clip,width=7.5cm]{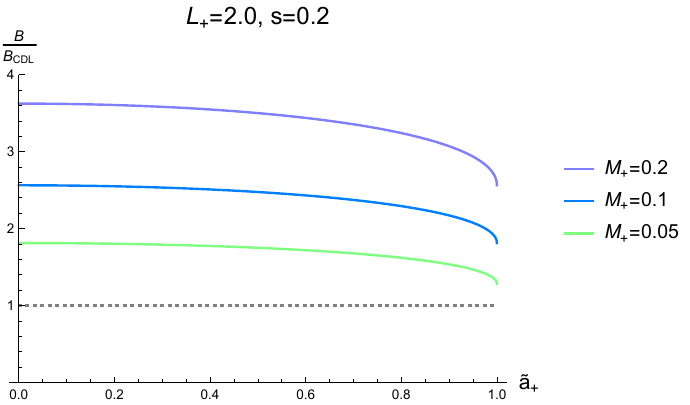}
      \end{minipage} &
      %---- 2番目の図 --------------------------
      \begin{minipage}[t]{0.45\hsize}
        \includegraphics[clip,width=7.5cm]{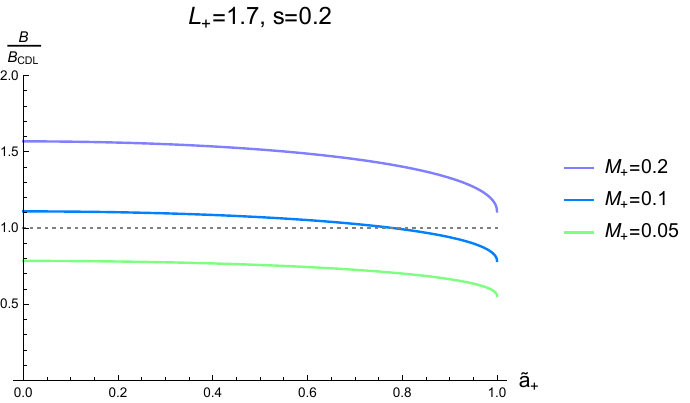}
      \end{minipage}
      %---- 図はここまで ----------------------
    \end{tabular}
     \caption{The $\tilde{a}_{+}$ dependence of $\mathcal{B}$ of static shells} \label{fig:Js02}
  \end{figure}

Motivated by the results that static shells dominate even 
for $\tilde{a}_{+}\neq0$, 
we consider the $\tilde{a}_{+}$ dependence of 
the decay rate of the static shell. 
Let us define $M_{s}$ as the value of $M_-$ for the static shell case.
Then 
$M_{s}$ 
decreases with $\tilde{a}_{+}$ (Fig.~\ref{fig:stm}), as expected from Fig.~\ref{fig:Jns}. 
The $\tilde a_+$ dependence of $\mathcal{B}$ for the static shell case is shown in Fig.~\ref{fig:Js02}, 
and we 
see that increasing $\tilde{a}_{+}$ raises the decay rate.

\section{Summary and Discussion}
\label{sec6}

In this paper, we have analysed vacuum decay in rotating BTZ BH spacetimes. 
In the computation, we have used the thin shell approximation and the Israel's junction condition to 
see the motion of the bubble wall.

%This result differs from the results for four-dimensional spacetimes shown in Refs.~\cite{Gregory:2013hja,Burda:2015yfa}. 
%Although the reason for this difference is not clear, it could be a non-trivial dependence on the number of dimensions and/or 
%the consequence of the mass gap specific to the BTZ BH solution\footnote{The 3-dimensional AdS spacetime corresponds to 
%$M=-\frac{1}{8}$, $J=0$ in the BTZ metric, and it is not realized in a continuous limit of the BTZ spacetime. }. 
%To see the dimensional dependence, 
%analyses in higher dimensional BH spacetimes, such as the five-dimensional Myers-Perry spacetime~\cite{Myers:1986un},  
%maybe helpful. 

As a result, we found that the decay rate can be lower than that of the Coleman De Luccia Instanton. 
We found that the decay rate is a decreasing function of $M_-$ and takes the maximum value at the lowest possible value of $M_-$, 
which is the case of the static shell. 
This result is partially consistent with results in Refs.~\cite{Gregory:2013hja,Burda:2015yfa}. 
In these previous works, however, in relatively 
small $M_{+}$ cases,  
bubble nucleation with no BH remnant 
dominates the probability.   
This difference stems from the difference in the change of the BH mass through the bubble nucleation.
In the previous works, 
the BH mass can either increase or decrease with the nucleation,
depending on the value of $M_+$. 
In particular, the black hole mass decreases 
for the small $M_+$ region and there is no static shell configuration for a
sufficiently small $M_+$ in the cases treated in Refs.~\cite{Gregory:2013hja,Burda:2015yfa}. 
In our 
cases, on the other hand, 
the BH mass must increase through bubble nucleation, and 
we can always have a static shell configuration~(see App.~\ref{secD} for the non-existence of the transition to 
a spacetime without a BH remnant).

We also found that the decay rate is a monotonic increasing function of $\tilde{a}_{+}$. 
That is, 
the angular momentum of the BTZ BH promotes the vacuum decay. 
On the other hand, the previous work on the Kerr spacetime \cite{Oshita:2019jan} stated that the angular momentum
suppresses the vacuum decay rate. 
Our results cannot be directly compared with those in Ref.~\cite{Oshita:2019jan} because of the 
totally different background settings and the 
assumptions
\footnote{
The authors in Ref.~\cite{Oshita:2019jan} focus on the cases 
$M_{+}=M_{-}, a_{+}=a_{-}, a^2_{\pm}/l^2 \ll 1$, where the 1st junction conditions are automatically satisfied
without angular dependencee. 
Then they introduce anisotropic tension to the stress energy tensor of the shell 
so that the 2nd junction condition can be satisfied. 
The form of the energy momentum tensor is clearly different from the isotropic tension adopted in our setting. 
We also note that 
the use of the Kerr-AdS geometry could be a nontrivial assumption 
because of the lack of the Birkhoff's theorem 
without spherical symmetry. 
}.
The analyses in the background Kerr spacetime are much harder than the analyses given here because of the lack of symmetry. 
Obviously, more investigations about the non-spherical bubble nucleation process would be needed to understand the effects of the angular momentum.

In this work, focusing on vacuum decay in the BTZ spacetime, 
we have found several non-trivial results which cannot be expected from the previous works. 
However, we do not have a clear understanding about the physical interpretation of these results, and more multidirectional study would be needed.

%%%%%%%%%%%%%%%%%%%%%%%%%%%%%%%%%%%%%%%%%%%%%%%%%%%%%%%%%%%%%%%%
\section*{Acknowledgements}
%%%%%%%%%%%%%%%%%%%%%%%%%%%%%%%%%%%%%%%%%%%%%%%%%%%%%%%%%%%%%%%%
This work was supported by JSPS KAKENHI Grant
Numbers JP19H01895~(C.Y.), JP20H05850~(C.Y.) and JP20H05853~(C.Y.).

\appendix

\section{Three-Dimensional Coleman De Luccia Instanton}
\label{secA}

In this appendix, we discuss vacuum decay in the pure AdS spacetime. 
The line element in the three-dimensional AdS spacetime is given by
\begin{align}
ds^2&=-h(r)dt^2+\frac{1}{h(r)}dr^2+r^2d\phi^2, \label{eq:ds'} 
\end{align}
where 
\begin{align}
h(r)&=1+\frac{r^2}{l^2}. 
\end{align}

As in the BTZ case, we set 
the shell   
$\mathcal{W}:r-R(t_{E})=0$ 
and consider the junction conditions. 
From the $(\phi,\phi)$ component of the 2nd condition, we get
\begin{align}
-\frac{1}{2}\dot{R}^2&:=U(R)=\frac{1}{2}\bar{\sigma}^2R^2-\frac{1}{2}\bar{h}+\frac{(\Delta h)^2}{32\bar{\sigma}^2R^2} , 
\end{align}
and by using the form of $h(r)$, we rewrite $U(R)$ as
\begin{align}
U(R)=\frac{1}{2}R^2\left[\left\{\bar{\sigma}-\frac{1}{4\bar{\sigma}}\left(\frac{1}{l^2_{+}}-\frac{1}{l^2_{-}}\right)\right\}^2-\frac{1}{l^2_{-}}\right]-\frac{1}{2} .
\end{align}
This is 
the potential that governs 
the motion of the Euclidean shell. The functional form of $U(R)$ is depicted in Fig. 13 for  $L_{+} = 2.0$, and $s = 0.2$.  
\begin{figure}[H]
\begin{center}
  \includegraphics[clip,width=6cm]{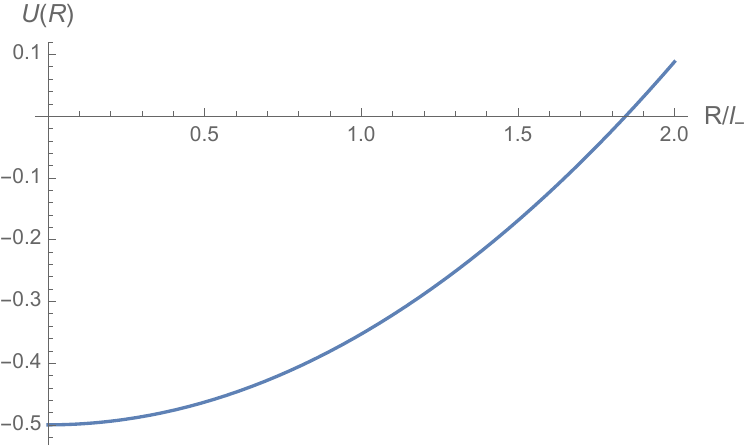}
  \end{center}
  \caption{Example of the potential of CDL instanton ($L_{+}=2.0, s=0.2$)}
\label{fig:CDL}
\end{figure}

We estimate the decay rate using the Euclidean action. 
Similarly to the BTZ case, we obtain the unit normal vectors $\tilde{u}^{\mu}_{\pm}$ and $\tilde{n}_{\pm\mu}$ as follows:
\begin{align}
\tilde{u}^{\mu}_{\pm}&=\left(\sqrt{g_{E\pm}^{tt}}, 0, 0\right), \\
\tilde{n}_{\pm\mu}&=(\mp\dot{R}, \pm\dot{t}_{E},0) .
\end{align}
Then we get
\begin{align}
\left.\tilde{n}_{\pm\mu}\tilde{u}^{\nu}_{\pm}\nabla_{\nu}\tilde{u}^{\mu}_{\pm}\right|_{\mathcal W}&=\left.\left(\mp\dot{R}\tilde{u}^{t_{E}}\nabla_{t_E}\tilde{u}^{t_{E}}\pm\dot{t}_{E}\tilde{u}^{t_{E}}\nabla_{t_E}\tilde{u}^{r}\right)\right|_{\mathcal W} \nonumber \\
&=\left.\mp\dot{t}_{E\pm}\left(-\frac{R}{l_{\pm}^2}\right)\right|_{\mathcal W},
\end{align}
where we used
\begin{align}
\Gamma^{r}_{{t}_{E}{t}_{E}}&=-h(r)\frac{R}{l^2}.
\end{align}
Then we get the bulk action as follows,
\begin{align}
S_{\mathcal{M}_{+}}+S_{\mathcal{M}_{-}}&=-2\int_{\mathcal{W}}d^2xR\sigma-\frac{1}{8\pi}\int_{\mathcal{W}}d^2xR\left(\frac{R}{l^2_{+}}\dot{t}_{E+}-\frac{R}{l^2_{-}}\dot{t}_{E-}\right) .
\end{align}
Because there is no horizon, $S_{\mathcal{H}}=0$, and we have the same $S_{\mathcal{W}}$ as the BTZ case. 
Finally, we obtain the following expression of the exponential factor: 
\begin{align}
\mathcal{B}_{CDL}&=S_{E} \nonumber \\
&=-\int_{\mathcal{W}}d^2xR\sigma-\frac{1}{8\pi}\int_{\mathcal{W}}d^2xR\left(\frac{R}{l^2_{+}}\dot{t}_{E+}-\frac{R}{l^2_{-}}\dot{t}_{E-}\right)\nonumber \\
&=\frac{1}{4}\oint d\tau\left(\dot{t}_{E+}-\dot{t}_{E-}\right) \nonumber  \\
&=\frac{1}{2}\left(L_{+}\sinh^{-1}\frac{4L_{+}s}{\sqrt{1+L_{+}^4(1-4s^2)^2-2L_{+}^2(1+4s^2)}}- 
   \sinh^{-1}\frac{4L_{+}^2s}{\sqrt{1+L_{+}^4(1-4s^2)^2-2L_{+}^2(1+4L_{+}s^2)}}\right) , \label{eq:CDL}
\end{align}
where
we used
\begin{align}
R\sigma=-\frac{1}{8\pi}[h\dot{t}_{E}]_{\pm}, 
\end{align}
that is derived from the $(\phi,\phi)$ component of Eq.~\eqref{eq:2nd'}.

\section{Static Shell}
\label{secB}

Here, we consider the static shell as a special case of the bounces. 
We can get this special solution by requiring $\dot{R}=0=\partial_{t_{E}}R$. 
In this case, we have
\begin{align} 
n_{\mu}&=\left(0,\frac{1}{\sqrt{f_{E}(r)}},0\right).
\end{align}
From the normalization, we have $\frac{1}{\sqrt{f_{E}(r)}}=\dot{t}_{E}$, and
\begin{align} 
n_{\mu}&=\left(0,\dot{t}_{E},0\right), \\
n^{\mu}&=\left(0,\frac{1}{\dot{t}_{E}},0\right).
\end{align}
Using this unit normal vector, the $(\phi,\phi)$ component of the extrinsic curvature is calculated as
\begin{align} 
K_{\phi\phi}=\sqrt{f_{E}(r)},
\end{align}
and from Eq.~\eqref{eq:2nd'}, we have 
the following equation:
\begin{align} 
\left.\left[\sqrt{f_{E+}(r)}-\sqrt{f_{E-}(r)}\right]\right|_{r=R}=-2\bar{\sigma}R. \label{eq:phiphi}
\end{align}
This equation is equivalent to $V(R)=0$. 
The $(\tau,\tau)$ component of Eq.~\eqref{eq:2nd'} is
\begin{align} 
\left.\left[\frac{\partial_{r}f_{E+}(r)}{2\sqrt{f_{E+}(r)}}-\frac{\partial_{r}
f_{E-}(r)}{2\sqrt{f_{E-}(r)}}\right]\right|_{r=R}=-2\bar{\sigma} ,
\end{align}
which is equivalent to $R$ derivative of Eq.~\eqref{eq:phiphi}, $V'(R)=0$. 

Next, we compute $\mathcal{B}$ for the static case. 
The unit normal vectors are written as
\begin{align}
\tilde{u}^{\mu}_{\pm}&=\left(\sqrt{g_{E\pm}^{tt}},0,\frac{g_{E\pm}^{t\phi}}{\sqrt{g_{E\pm}^{tt}}}\right), \\
\tilde{n}_{\pm\mu}&=(0,\pm\dot{t}_{E},0) ,
\end{align}
so that 
$\tilde{u}^{\mu}_{\pm}\tilde{n}_{\pm\mu}=0$. 
Hence, we get
\begin{align}
\frac{1}{8\pi}\int_{\mathcal{W}}d^2x\sqrt{h_{E}}(n_{+\mu}\tilde{u}^{\nu}_{+}\nabla_{\nu}\tilde{u}^{\mu}_{+}-n_{-\mu}\tilde{u}^{\nu}_{-}\nabla_{\nu}\tilde{u}^{\mu}_{-})
&=-\frac{1}{8\pi}\int_{\mathcal{W}}d^2x\sqrt{h_{E}}(\tilde{u}^{\mu}_{+}\tilde{u}^{\nu}_{+}\nabla_{\nu}n_{+\mu}-\tilde{u}^{\mu}_{-}\tilde{u}^{\nu}_{-}\nabla_{\nu}n_{-\mu}) \nonumber \\
&=-\frac{1}{8\pi}\int_{\mathcal{W}}d^2x\sqrt{h_{E}}(K_{+\tau\tau}-K_{-\tau\tau}).
\end{align}
Here, from Eq.~\eqref{eq:2nd'},
\begin{align}
K_{E+\tau\tau}-K_{E-\tau\tau}&=-8\pi\sigma
 \end{align}
is satisfied, and we can write the bulk action as
\begin{align}
S_{\mathcal{M}_{+}}+S_{\mathcal{M}_{-}}&=\frac{1}{8\pi}\int_{\mathcal{W}}d^2x\sqrt{h_{E}}(-16\pi\sigma)-\frac{1}{8\pi}\int_{\mathcal{W}}d^2x\sqrt{h_{E}}(-8\pi\sigma) \nonumber\\
&=-\int_{\mathcal{W}}d^2x\sqrt{h_{E}}\sigma .
\end{align}
This is the same as $S_{\mathcal{W}}$ in Eq.~\eqref{eq:Sw} with the opposite sign, 
and they cancel out. 
Eventually, the only nonzero contribution to $S_{E}$ is from $S_{\mathcal{H}_{-}}$.

The contribution from the false vacuum can be written as
\begin{align}
S_{E0}=-\frac{A_{\mathcal{H}_{+}}}{4},
\end{align}
so that we can write $\mathcal{B}$ as
\begin{align}
\mathcal{B}=\frac{1}{4}(A_{\mathcal{H}_{+}}-A_{\mathcal{H}_{-}}),
\end{align}
which indicates that the decay rate depends only on the BH horizon areas. 
We note that this simple result comes from 
the relation $\tilde{u}^{\mu}_{\pm}\tilde{n}_{\pm\mu}=0$, originated from the staticness of the shell.

\section{Horizon Area}
\label{secC}

In this appendix, 
we show that the horizon area decreases through the nucleation under 
the conditions in Sec.~\ref{sec3}.
Since the spatial section of the horizon is a one-dimensional object,
its change is proportional to the difference of the radius, that is, 
\begin{align} 
\frac{1}{l_{-}}\Delta r_{\mathcal{H}}&:=\frac{1}{l_{-}}(r_{\mathcal{H}_{+}}-r_{\mathcal{H}_{-}}) \nonumber \\
&=\sqrt{4M_{+}L_{+}^2+4M_{+}L^2_{+}\sqrt{1-\tilde{a}_{+}^2}}-\sqrt{4M_{-}+4\sqrt{M_{-}^2-M^2_{+}L^2_{+}\tilde{a}_{+}^2}} .\label{eq:delr}
\end{align}
From Eq.~\eqref{eq:delr}, for $\Delta r_{\mathcal{H}}$ to be positive, $M_{-}$ must satisfy the condition
\begin{align} 
M_{-}<M_{r}:=\frac{M_{+}}{2}\left[L_{+}^2+1+(L_{+}^2-1)\sqrt{1-\tilde{a}_{+}^2}\right].
\end{align}
By comparing $M_{r}$ with $M_{T}$, which gives
the upper bound of $M_{-}$ from the condition $\dot{t}_{E\pm}>0$, we find 
\begin{align} 
M_{r}-M_{T}=2M_{+}L^2_{+}s^2\left[1+\sqrt{1-\tilde{a}_{+}^2}\right]>0.
\end{align}
Therefore, as long as we focus on the nucleation 
with
$M_{-}<M_{T}$, 
the horizon area 
decrease. 
Meanwhile, for 
%the 
a
shrinking solution after the nucleation, 
the horizon area increases through the classical process of the shell accretion.

\section{Possibility of Other Bounce Solutions}
\label{secD}

\begin{figure}[H]
\begin{center}
  \includegraphics[clip,width=6cm]{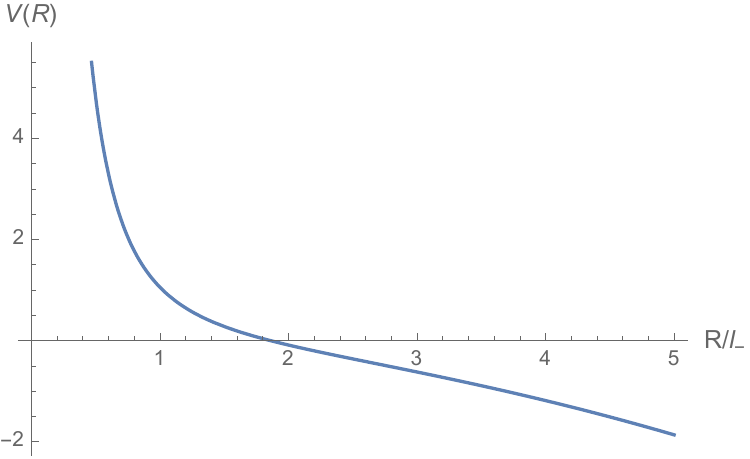}
  \end{center}
  \caption{The potential form for $L_{+}=2.0$, $s=0.3$, $\tilde{a}_{+}=0$, $M_{+}=0.1$ and $M_{-}=0.336$}
\label{fig:negaA}
\end{figure}

Let us consider the possibility of other bounce solutions. 
First, we consider the case $A<0$, $C>0$ and $\dot{t}_{E\pm}>0$. 
The potential form is given like Fig.~\ref{fig:negaA}. 
In this case, although we may consider Euclidean shell motion in the region $V(R)<0$, 
after the nucleation, the Lorentzian shell shrinks. 
Thus we do not consider this case.  

\begin{figure}[H]
\begin{center}
  \includegraphics[clip,width=6cm]{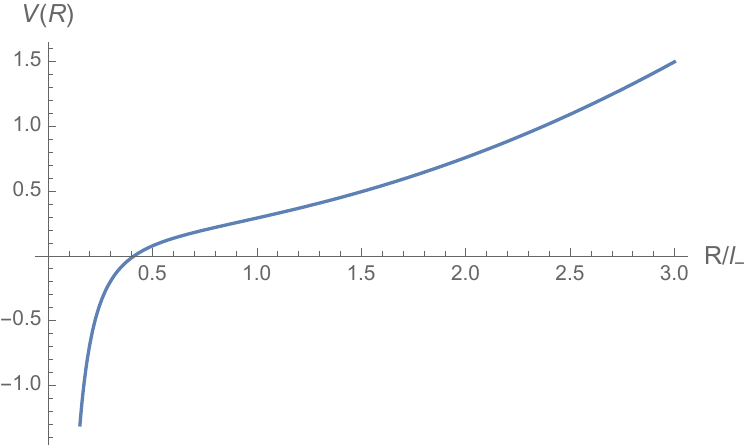}
  \end{center}
  \caption{The potential form for $L_{+}=2.0$, $s=0.2$, $\tilde{a}_{+}=0$, $M_{+}=0.1$ and $M_{-}=0.13$)}
\label{fig:negaC}
\end{figure}

Next, we consider 
the case $A>0$, $C<0$ and $\dot{t}_{E\pm}>0$. 
In this case, the potential form is given 
like Fig.~\ref{fig:negaC}. 
About the motion of the Euclidean bounce, 
one might expect that a shell emerges from the one point $R=0$, and 
expands until the turn around point at $V(R)=0$, 
similarly to the CDL case. 
In the BH case, however, the point $R=0$ corresponds to
the singularity, and the bounce solution is also singular at $R=0$.  
Therefore we do not consider this case. 

Finally, let us consider the transition between a BTZ and the pure AdS spacetime, which is given by $M=-\frac{1}{8}$ and $J=0$. 
There is two 
possibilities 
for this sort of nucleation : 
the transition between the BTZ spacetime and the pure AdS spacetime which corresponds to true/false vacuum.

First, we consider the former case, namely, $M_-=-\frac{1}{8}$ and $M_+>0$. 
By using Eq.~\eqref{eq:sfin}, 
we obtain the lower bound for $B$:
\begin{align} 
B&>4(M_{+}+M_{-})-\frac{L_{+}^{-2}-1}{\frac{1}{4}L_{+}^{-2}-\frac{1}{2}L_{+}^{-1}+\frac{1}{4}}(M_{+}-M_{-}) \nonumber \\
&=\frac{8(M_{+}L_{+}-M_{-})}{L_{+}-1}.
\end{align}
With $M_{-}=-\frac{1}{8}$, we see
\begin{align} 
B&>\frac{8M_{+}L_{+}+1}{L_{+}-1}>0, 
\end{align}
and $B$ cannot be negative and we cannot make a bounce solution. 
Next, we consider the the case $M_+=-\frac{1}{8}$ and $M_->0$. 
By comparing $M_{D}$ and $M_{T}$, we obtain
\begin{align} 
M_{D}-M_{T}&=\frac{M_{+}}{2}\sqrt{1-\tilde{a}_{+}^2}\left[1+L_{+}^2(-1+4s^2)+\sqrt{1+L_{+}^4(1-4s^2)^2-2L_{+}^2 (1 + 4 s^2)}\right],
\end{align}
and we see $1+L_{+}^2(-1+4s^2)<0$ from Eq.~\eqref{eq:sfin}. 
By noting that 
\begin{align} 
&\left[1+L_{+}^2\left(-1+4s^2\right)\right]^2-\left[1+L_{+}^4\left(1-4s^2\right)^2-2L_{+}^2\left(1+4s^2\right)\right]=16L_{+}^2s^2>0, \nonumber \\
&1+L_{+}^2(-1+4s^2)-\sqrt{1+L_{+}^4(1-4s^2)^2-2L_{+}^2 (1 + 4 s^2)}<0,
\end{align}
we obtain $1+L_{+}^2(-1+4s^2)+\sqrt{1+L_{+}^4(1-4s^2)^2-2L_{+}^2 (1 + 4 s^2)}<0$. 
From that, we see $M_{T}<M_{D}$ when $M_{+}<0$, so there is no parameter region of $M_{-}$ such that $M_{D}<M_{-}<M_{T}$. 
Instead of $M_{D}$, let us adopt $M'_{D}$ as the bound for $M_{-}$. 
In this case,
\begin{align} 
M_{C}-M'_{D}&=\frac{M_{+}}{2}\left[1+4\tilde{a}_{+}L_{+}s+L_{+}^2(-1+4s^2)+\sqrt{(1-\tilde{a}_{+}^2)\left\{1+L_{+}^4(1-4s^2)^2-2L_{+}^2(1+4s^2)\right\}}\right]
\end{align}
and $1+4\tilde{a}_{+}L_{+}s+L_{+}^2(-1+4s^2)<0$ follows from Eq.~\eqref{eq:sfin}. 
By noting that 
\begin{align} 
&\left[1+4\tilde{a}_{+}L_{+}s+L_{+}^2(-1+4s^2)\right]^2-\left[(1-\tilde{a}_{+}^2)\left\{1+L_{+}^4(1-4s^2)^2-2L_{+}^2(1+4s^2)\right\}\right] \nonumber \\
&=\left[\tilde{a}_{+}+4L_{+}s+\tilde{a}_{+}L_{+}^2(-1+4s^2)\right]^2>0, \nonumber \\
&1+4\tilde{a}_{+}L_{+}s+L_{+}^2(-1+4s^2)-\sqrt{(1-\tilde{a}_{+}^2)\left\{1+L_{+}^4(1-4s^2)^2-2L_{+}^2(1+4s^2)\right\}}<0,
\end{align}
we obtain $M'_{D}<M_{C}$ for $M_{+}<0$. 
Then there is no parameter region of $M_{-}$ such that $M_{C}<M_{-}<M'_{D}$.  
Therefore there is not the case in which the potential form is given like Fig.~\ref{fig:potex}. 

From the above results, we conclude that the cases we considered in the main part of this paper are the only possible %and 
physical bounces in BTZ spacetimes.

\bibliography{hoge} 
\bibliographystyle{unsrt}

\end{document}